# Time series classification based on fractal properties

1st Vitalii Bulakh
*dept. Applying mathematics*
Kharkiv National University of Radioelectronics
Kharkiv, Ukraine
bulakhvitalii@gmail.com

2nd Lyudmyla Kirichenko
*dept. Applying mathematics*
Kharkiv National University of Radioelectronics
Kharkiv, Ukraine
lyudmyla.kirichenko@nure.ua

3rd Tamara Radivilova
*dept. of Infocommunication Engineering*
Kharkiv National University of Radioelectronics
Kharkiv, Ukraine
tamara.radivilova@nure.ua

*Abstract* – **The article considers classification task of fractal time series by the meta algorithms based on decision trees. Binomial multiplicative stochastic cascades are used as input time series. Comparative analysis of the classification approaches based on different features is carried out. The results indicate the advantage of the machine learning methods over the traditional estimating the degree of self-similarity.**

*Keywords – multifractal time series, binomial stochastic cascade, classification of time series, Hurst exponent, Random Forest*

## I. Introduction

Many complex processes have a fractal structure and their dynamics are represented by time series possessing fractal properties. Such processes include various information processes in communication networks, including hacker attacks.

Distributed Denial of Service (DdoS) attack is hacker attack on the computer system in order to force it to failure that is creation of such conditions at which conscientious system users can not access the provided system resources, or this access is difficult. Failure of the system can be a step towards mastering it. Currently, DoS and DDoS-attacks are the most popular, since they allow to force almost any system to failure without leaving legally significant evidence. One solution to the problem of early intrusion detection is the development of a classifier that would determine the probability that incoming traffic contains an attack.

Recent researches show that network traffic has fractal properties and one of the characteristic features of intrusion is changing of fractal characteristics in the traffic containing attack. In [1-3] a number of methods of detection based on changing traffic fractal structure containing attacks. An important step in the intrusion detection methods is classification of the test traffic on the fractal properties, in particular, range of values of the Hurst parameter.

In many cases, the problems of recognizing and classifying fractal series take place. Most often, such tasks are solved by estimating and analyzing fractal characteristics [4-7]. However, in recent years, there has been a growing interest in machine learning methods to analyze and classify fractal series [8-10]. The aim of this paper is a comparative analysis of the classification of fractal stochastic time series performed by meta-algorithms using decision tree methods.

## II. Characteristics of Self-Similar and Multifractal Processes

The self-similarity of random processes is to preserve statistical characteristics when changing the time scale. A stochastic process $X(t)$ is self-similar with a parameter $H$ if the process $a^{-H}X(at)$ is described by the same finite-dimensional distribution laws as $X(t)$. The parameter $H$, $0 < H < 1$, called the Hurst exponent, represents the self-similarity degree. Along with this property, the Hurst exponent characterizes the measure of the long-term dependence of the stochastic process. The moments of the self-similar random process satisfy the following scaling relation

$$E\left[|X(t)|^q\right] \propto t^{qH}. \qquad (1)$$

Multifractal random processes are inhomogeneous fractal ones and have more flexible scaling laws for moment characteristics [11]:

$$E\left[|X(t)|^q\right] \propto t^{qh(q)}, \qquad (2)$$

where $h(q)$ is generalized Hurst exponent, in the general case, a nonlinear function for which the value $h(q)$ at $q=2$ coincides with the value of Hurst exponent $H$. For time series that correspond to a monofractal process, the generalized Hurst exponent $h(q)$ is constant: $h(q) = H$.

## III. Estimating of Fractal Characteristics by Time Series

There are many methods for estimating the fractal characteristics by time series. One of the most popular is the method of multifractal detrended fluctuation analysis (MFDFA) [12]. MFDFA is powerful tool for statistical processing of time-dependent processes.

According to the MFDFA method, the fluctuation function $F^2(\tau) = \frac{1}{\tau}\sum_{t=1}^{\tau}(y(t) - Y_m(t))^2$ is calculated for each segment of length $\tau$, where $y(t)$ is input cumulative time series; function $Y_m(t)$ is local *m*-polynomial trend within the given segment. If the investigated series $y(t)$ has a long-term dependence the averaged function $F(\tau)$ has scaling $F(\tau) \propto \tau^H$.

In the study of multifractal properties the dependence of the fluctuation function $F_q(\tau)$ of a parameter $q$ is considered:

$$F_q(\tau) = \left\{ \frac{1}{N} \sum_{i=1}^{N} [F^2(\tau)]^{\frac{q}{2}} \right\}^{\frac{1}{q}}.$$

If the investigated series is multifractal and has a long-term dependence, the fluctuation function $F_q(\tau)$ has scaling $F_q(\tau) \propto \tau^{h(q)}$. Estimate of the Hurst exponent $H$ can be represented by confidence interval within which the true value $H$ is found [13-15]:

$$\hat{H} + \Delta - t_\alpha S < H < \hat{H} + \Delta + t_\alpha S, \quad (3)$$

where $\hat{H} = \hat{H}(N)$ is obtained evaluate of $H$; $N$ is the time series length; $\Delta = \Delta(N)$ is the calculated mean bias of the estimate, $S = S(N)$ is the calculated standard deviation; $\alpha$ is required significance level; $t_\alpha$ is the quantile of the simple normal distribution. Values $\Delta$ and $S$ can be obtained numerically for different lengths of time series.

IV. CLASSIFICATION OF TIME SERIES BY DECISION TREE METHODS

The decision tree method is applicable to solving classification problems arising in various fields and it is considered one of the most effective. It consists in the process of dividing the original data set into groups, until homogeneous subsets are obtained. The set of rules that give such a partition allows to make a conclusion, produced based on an evaluation of some input attributes for new data.

The algorithm of learning a tree acts on the principle of recursive partitioning. Partitioning the data set is based on the use of the most suitable for this feature. In the tree a corresponding decision node is created, and the process continues recursively until the stop criterion is fulfilled.

The models that adapt their state in the learning process in accordance with the training set, such as decision trees, are unstable: even a small change in the training set may cause to significant changes in the structure of the tree. In other words, by making even minor changes in the training data, we will always receive a different model. But the original and modified models will operate approximately the same way and with comparable accuracy: minor changes in the training data will not lead to a change in the basic regularities. In this case, it is expedient to use ensembles of models. The ensemble of models as a whole can be considered as a composite model consisting of separate basic models. The components of the ensemble can be both the same type and different.

One of the first and most famous types of ensembles is the bagging method based on the statistical method of bootstrap aggregating [16]. Bootstrap aggregating is a computer method for studying the distributions of probability distribution statistics, obtained by multiple generation of samples based on a single sample.

Bagging is a classification technique where all elementary classifiers are trained and operate independently of each other. The idea is that the classifiers do not correct each other's mistakes, but compensate them for voting. The basis of the bagging method is the classification technology, called "perturbation and combination". Perturbation is understood as the introducing of some random changes in training data and the construction of several alternative models on the modified data, followed by a result combination. From a single training set several samples containing the same number of objects are extracted by sampling. Each of the received samples is used to train one of the ensemble models. If the ensemble is built on the basis of models of various type, then each type has its own learning algorithm.

To obtain the result of the work of models ensemble the following combination methods are usually used: voting (a class that was chosen by a simple majority of ensemble models) or averaging which can be defined as the simple average of the outputs of all models (if weighted averaging is performed, then the model outputs are multiplied by the corresponding weights). The effectiveness of bagging is achieved due to the fact that the basic algorithms, trained in different subsamples, are obtained quite different and their mistakes are mutually compensated for in the voting process and also because the emission objects may not be included in some training subsamples.

Random forest is also a method of bagging, but unlike its main version, the random forest has several features: 1) it uses within itself an ensemble of only regression or classifying decision trees; 2) in the sampling algorithm, in addition to random selecting training objects random selection of features is also carried (usually the new number of features is equal to the square root of the total number); 3) for each subsample, the decision tree is built up to the full completion of the training objects and it is not subjected to post pruning. [17]

V. INPUT DATA

Numerous studies of processes in information networks have shown that network traffic has the multifractal properties. One of the frequently used models of the multifractal process is the conservative stochastic binomial multiplicative cascade [18]. In its construction, the initial time interval is divided into two equal intervals, to which the weighting factors $w_1$ and $1 - w_1$ are assigned. The weighting coefficients are the independent values of some given random variable. If we choose a random variable defined over the interval $[0, 1]$, then the sum of the coefficients at each iteration is equal to unity. On the second iteration, two new independent random values $w_2$ and $w_3$ are added. We obtain four intervals with weight coefficients $w_1 w_2$, $w_1(1-w_2)$, $(1-w_1)w_3$ and $(1-w_1)(1-w_3)$. With the number of iterations $n \to \infty$, we arrive at the limiting measure, which is an inhomogeneous fractal set. For each iteration $n$, $n \gg 1$, we have a time series (cascade) of length $2^n$ with multifractal properties.

The fractal characteristics of stochastic multiplicative cascade obtained using beta distribution random variable $Beta(\alpha, \beta)$ are completely determined by the parameters $\alpha, \beta > 0$ [19]. Fig. 1 shows typical cascades with different Hurst exponent: the cascade with $H = 0.7$, is on the left, on the right is the cascade with values $H = 0.9$.

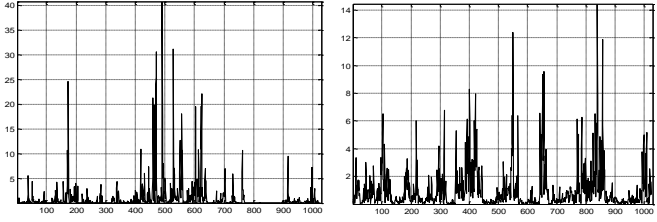

Fig. 1. Multifractal cascades: on the left $H = 0.7$, on the right $H = 0.9$

## VI. EXPERIMENT AND RESULTS

Fractal time series for classification were obtained by generating stochastic binomial cascades on the basis of a symmetric beta distribution. In this case, building the cascade process allows to obtain multifractal time series with the Hurst exponent in the range $H \in (0.5, 1)$ and they can be split into several classes according to the Hurst exponent values.

Each class is a set of generated time series with a Hurst exponent, which lies in a specified range of values. To generate cascades within each class values $H$ are chosen using a uniform distribution. To form series classes the ranges of Hurst exponent are changed in the interval $(0.5, 1)$ in increments of 0.05. The minimum and maximum values of the $H$ were chosen at 0.51 and 0.99, respectively. Thus, the training of models was carried out on 11 classes, where $H \in \{[0.51, 0.525), [0.525, 0.575), [0.575, 0.625), \ldots, [0.975, 0.99]\}$.

A preliminary comparative analysis of the classification of fractal time series by meta-algorithms using decision tree methods was carried out. The results showed that the best results are given by the method of Bagging and Random Forest which use regression trees. Therefore, to carry out the classification using the methods of Machine Learning, the Random Forest method with regression trees was selected.

In the work, three different approaches to determine the time series belonging to one of the classes were considered. In the first approach, the classification was carried out directly from the time series values by the Random Forest method, where the object was the cascade time series, and the features were the values of this series. When using decision tree regression the result is the probability of cascade series matching to a given class The probabilities are calculated by the formula: $P_i = 1 - |m_i - C|$, where $m_i$ is the regression result for the $i$-th example, $C$ is theoretically known class number.

The second approach to the classification of multifractal cascades is also based on machine learning, but in this case, the statistical and multifractal characteristics obtained from time series were used for the classification: standard deviation, maximum value and median of series, mean and standard deviation of the generalized Hurst exponent $h(q)$, the values $h(1)$, $h(2) = H$, and the range $\Delta h = h(0.1) - h(5)$. In this case, the objects were cascade series, and the features were the estimates of characteristics calculated for each cascade.

The third way of classification is based on direct estimation of the Hurst exponent by time series and determination of the confidence interval in accordance with (3). The choice of the class was defined as the probability of finding the true value of the Hurst exponent within the confidence interval in relation to the class width.

To build decision tree models, Python with libraries that implement machine learning methods was used. The training of models for each class was conducted on 500 examples of time series and tested on 50 test cases. In the case of using confidence intervals, the classification was carried out for the test sample. The classification was performed for time series with different lengths, but for compare the results, the main attention was paid to the series of length 512 and 4096 values.

As a classification result, the histograms of the probabilities of class determination for each range of the Hurst exponent were obtained. Fig. 2 shows the histograms obtained for the class $0.725 \leq H < 0.775$ corresponding to the classification results for cascades of length 4096 values. Such distributions are typical for all cascade classes, including for series with greater number of values.

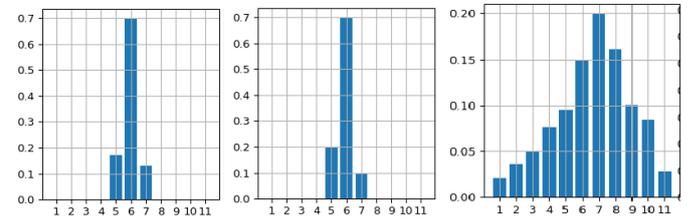

Fig. 2. Distribution of probabilities of determining the class number for different classification approaches: by time series values (left), by time series characteristics (middle), by estimation of Hurst exponent (right)

In many classification tasks, including detecting a discord or changing one of the system indicators, it is required to determine the probability of the object belonging to one of the two classes (for example, to detect that the server is being attacked). Therefore, in the work the classification of cascade series into 2 classes was carried out, for which the Hurst exponent was chosen in the ranges $0.51 \leq H < 0.7$ and $0.7 \leq H \leq 0.99$.

The table presents the average probability of class determination and time of training model depending on the classification method for both cases: for 11 and 2 time series classes. Thus, the obtained results indicate a great advantage of machine learning methods over traditional methods of estimating fractal characteristics when classifying time series by fractal properties. However, for training, a sufficient number of time series with known properties is necessary, which is often problematic. Obviously, an interesting and promising task is to develop a classification methods based on the simulated time series with given fractal and probabilistic properties for using them along with real data as a training sample.

TABLE I. AVERAGE PROBABILITY OF CLASS DETERMINATION

|  | Length | Time series values | | Time series characteristics | | Estimate H |
|---|---|---|---|---|---|---|
|  |  | P | time | P | time | P |
| *11 classes* | 512 | 0.75 | 15 min | 0.74 | 1 min | 0.19 |
|  | 4096 | 0.81 | 120 min | 0.75 | 1.5 min | 0.23 |
| *2 classes* | 512 | 0.97 | 5 sec | 0.96 | 15 sec | 0.75 |
|  | 4096 | 0.97 | 10 sec | 0.96 | 25 sec | 0.78 |

Classification performed on the basis of training by the time series values showed a slightly higher accuracy than on the basis fractal and statistical characteristics. However, it should be noted that in this case the number of features is determined by the series length. It requires learning anew when changing the length of the time series, moreover, the time of leaning increases nonlinearly with increasing length of the time series. The time of learning the model by characteristics weakly depend on the series length and it is substantially smaller for large series. It is worth emphasizing that in the second case, the possibility of using the learned model does not depend on the series length: with a small length of the series, only the accuracy of the classification can suffer because of less estimation accuracy of characteristics.

## VII. CONCLUSION

In this paper, a comparative analysis of the classification of multifractal stochastic time series using meta-algorithms based on decision trees has been performed. Binomial multiplicative stochastic cascades were used as input time series. Time series were split into several classes depending on their degree of self-similarity.

Three different approaches were used for the classification. In the first, the classification was carried out by the Random Forest method directly by the time series values. In the second, Random Forest was also used, but the features for classification were statistical and multifractal characteristics of the time series. In the third case, the probability of belonging to the class was determined by estimating the Hurst exponent by the time series.

The results of the classification showed the advantage of machine learning methods over the traditional method of estimating the Hurst exponent, especially with a short length of the time series.

The obtained results can be used for practical applications related to the classification or clustering of real time series with fractal properties, for example, detecting DDoS attacks.

In our future research we intend to concentrate on the study and classification of traffic realizations that includes DDoS attacks and without that ones and for various transmission protocols in order to develop a method of early intrusion detection based on the use machine learning methods.